\documentclass[12pt]{article}
\usepackage[margin=1in]{geometry}
\usepackage{doublespace}
\usepackage{graphicx}
\usepackage{amssymb, amsthm, amsmath}
\usepackage{bm}

\usepackage[algo2e,ruled,noline]{algorithm2e} 
\usepackage[numbers]{natbib}
\usepackage{float}
\usepackage{placeins}
\usepackage{hyperref}
\renewcommand{\url}[1]{\newline\normalfont{#1}}
\usepackage{titlesec}
\titleformat{\section}
  {\normalfont\bfseries}
  {\thesection}
  {1em}
  {}
  
\titleformat{\subsection}
  {\normalfont\bfseries\slshape}
  {\thesubsection}
  {1em}
  {}

\titleformat{\subsubsection}
  {\normalfont\slshape}
  {\thesubsubsection}
  {1em}
  {}

\begin{document}

\title{Selecting Diverse Models for Scientific Insight\protect}
\begin{center}
{\Large Selecting Diverse Models for Scientific Insight}\\\vspace{6pt}
{\large Laura J. Wendelberger, Brian J. Reich, and Alyson G. Wilson}\\
{\large Department of Statistics, North Carolina State University}\\
\end{center}


\begin{abstract}
\noindent Model selection often aims to choose a single model, assuming that the form of the model is correct. However, there may be multiple possible underlying explanatory patterns in a set of predictors that could explain a response. Model selection without regard for model uncertainty can fail to bring these patterns to light. We explore multi-model penalized regression (MMPR) to acknowledge model uncertainty in the context of penalized regression. We examine how different penalty settings can promote either shrinkage or sparsity of coefficients in separate models. The method is tuned to explicitly limit model similarity. A choice of penalty form that enforces variable selection is applied to predict stacking fault energy (SFE) from steel alloy composition. The aim is to identify multiple models with different subsets of covariates that explain a single type of response.
\vspace{12pt} \end{abstract}



\section{Introduction}\label{s:intro}
Analysts often identify a single best-fitting model to be used for subsequent prediction. However, there may be several models that fit well. When the goal of an analysis is to better understand the relationship between the covariates and the response, fitting a single model neglects the possibility of additional models with similar explanatory power. For example, comparing model fitting to route mapping, when we search Google Maps for a route, it often presents several route options; one may take back roads while another may take a more direct highway, and a third may take a toll road. Depending on your travel needs, you may prefer one route to the others, so it is beneficial to be aware of all of the options. If we consider model fitting in a similar way, it is valuable to identify a range of different models that relate the predictors to the response and reveal distinct representations of the data.

\subsection{A motivating example}
As motivation, we consider a materials science problem relating alloy composition to failure mode. Steel components are widely used in tools, construction, infrastructure, and more. Steel alloys are composed of iron (Fe) and carbon (C), with additional elements that determine the material properties of the alloy. Specifically, austenitic steels are non-magnetic stainless steel alloys characterized by relatively high levels of chromium (Cr) and nickel (Ni). Avoiding failure of components made from these steel alloys is important for safety and quality. 

Deformations are a known failure mechanism for steel components. At a microstructure level, a deviation in the order of the stacked atomic planes from the expected crystal structure is called a \textit{stacking fault}. The stacking fault energy (SFE) is the energy associated with this disorder, and it can be useful in predicting the mode of deformation in austenitic steels \citep{Allain2004}. Knowledge of the mode of deformation is important for satisfying design criteria for the component.

Historically, materials scientists have used theoretical and computational tools to model material properties. When these methods exist, they are sometimes poor representations of the processes because of unmodeled physics. For example, there are first-principles computational models for SFE that are based on quantum mechanics \citep{Vitos2006,Lu2011}. However, these models can be computationally costly and cannot capture all of the underlying physics. Even when sufficient physical understanding is available, including it can result in prohibitively slow computation speeds, making screening of large materials data sets impractical. While SFE cannot be experimentally measured, it can be inferred from transmission electron microscopy (TEM) or diffraction techniques. Typically, screening a large dataset of different compositions experimentally is infeasible. Consequently, materials scientists have begun to adopt data-driven approaches to modeling material properties \citep{Ramprasad2017}. Local linear models on small datasets from individual experiments have been used as a first statistical approach to relate composition to SFE.

\begin{figure}[h]
    \centering
    \includegraphics[width=0.7\linewidth]{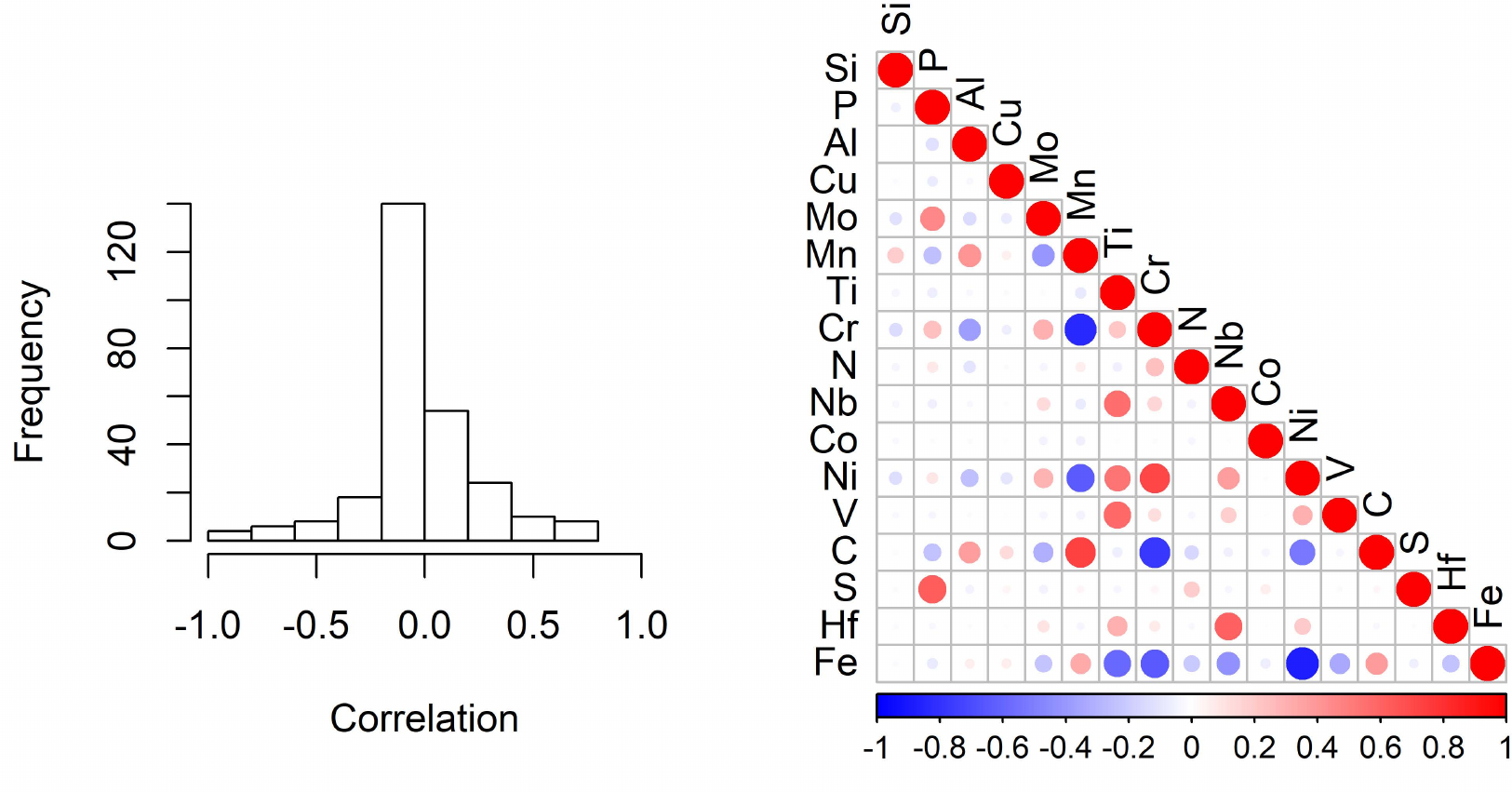}
    \caption{Histogram (left) and correlation plot (right) of observed covariance for SFE data.}
    \label{fig:cor_steel_comp}
\end{figure}

\cite{Chaudhary2017} collected data from many different studies to form a combined dataset with chemical composition and experimentally measured SFE for austenitic steels. The dataset, accessed on the Citrination platform ~\citep{citrine}, has $n=946$ observations of experimentally measured SFE for different compositions of austenitic steel. The composition of each observation is reported in terms of weight percentage for 17 different alloying elements. There are 10 observations where the composition of vanadium (V) is missing, but we assume it to be zero since the compositions of the other elements in these structures sum to 100. The observed correlation structure among these $p=17$ covariates (Figure~\ref{fig:cor_steel_comp}) shows that many of the covariates display little to no correlation with each other. 

\citeauthor{Chaudhary2017} apply a variety of machine learning techniques to predict SFE from steel composition, with more focus on prediction and less on model interpretability. An interpretable model reveals the extent to which the composition elements, or even groups of elements, in an alloy relate to the resulting stacking fault energy. This type of analysis is suitable for the SFE data because the aim is to understand why certain alloys have higher SFE rather than simply to predict SFE. In addition, correlation among the alloy components can make it difficult to capture all of the information in a single model. Multiple different models can reveal several potential relationships between alloying elements and SFE. Our goal is to identify several different possible explanatory models for SFE and to provide several possible interpretable explanations of the observed data.

\subsection{Literature review}\label{s:lit}
Broadly, model uncertainty refers to uncertainty regarding the form of the model. Very generally, model uncertainty may be captured by a model discrepancy term to, for example, capture the inadequacy of computer simulation to fully describe the data generating process behind experimental data \citep{Kennedy2001}. Since we restrict the models under consideration to linear models, the model uncertainty refers to the subset of variables to include in the model. In this paper, we consider model selection as a variable selection process, which allows us to include meaningful variables in a statistical model while excluding nonessential ones. 

Penalized regression methods are common variable selection tools that use different penalties to promote shrinkage \citep{Hastie2015,Hastie2017}. The least absolute shrinkage and selection operator (LASSO) technique is a frequently used approach for model selection because it obtains model sparsity by shrinking non-influential variable effects to exactly zero while maintaining computational efficiency \citep{Tibshirani1996,Zou2006}. In the presence of multicollinearity, however, the LASSO-selected values are unstable. Some extensions take into account prior knowledge about correlated covariates to include or exclude them from the model as a group via the fused LASSO \citep{Tibshirani2005} and the grouped LASSO \citep{Yuan2006}. Taking this a step further, the OSCAR method performs supervised grouping of important variables without a prior specification \citep{Bondell2008}. 

Existing penalized regression methods attempt to fit one true, best model. While straightforward, this approach ignores the concept of model uncertainty. The Random LASSO \citep{SijianWang2011}, developed in the context of highly correlated microarray data, incorporates model uncertainty by applying LASSO to bootstrapped samples of covariates in a multi-stage procedure and generating a weighted averaged model, automatically realizing grouped covariate inclusion/exclusion behavior.  Split regularized regression (SplitReg) \citep{Christidis2020} addresses model uncertainty by proposing a multi-model objective function to construct a set of diverse and sparse models. The model average of this ensemble has improved prediction capabilities. Tuning parameters are selected via cross validation, suggesting that weight is given to model fit, possibly at the cost of identifying diverse models.

Outside of penalized regression, Bayesian Model Averaging (BMA) was developed to account for model uncertainty using a Bayesian hierarchical framework \citep{Madigan1994,Raftery1995,Draper1995}. BMA considers multiple model configurations to calculate posterior distributions over both models and coefficients \citep{Clyde2004,Hoeting1999,Burnham2010}. An exhaustive BMA approach considers the fits models fit using the $2^p$ possible subsets of $p$ covariates and assigns a posterior probability to each. However, this method typically generates highly correlated models that are often slightly modified iterations of the best model. In addition, it may be necessary to sift through thousands of model candidates when performing BMA to identify a set of uncorrelated models.

Several machine learning techniques incorporate model uncertainty through the use of ensemble techniques to improve stability and reduce variance by incorporating information from multiple model fits \citep{Hastie2017}. Bootstrap aggregation (bagging) creates multiple models by fitting many subsets of the data and then averaging the output. These models can vary widely, especially with highly correlated predictors \citep{Breiman1996,Breiman1996a}. \cite{Ho1998} introduced the random subspace method, which generates trees using pseudorandom subsets of the available features, ultimately reducing the correlation between the trees. Random forests \citep{Breiman2001} combine the advantages of bagging and the random subspace method to produce both accurate estimation and a measure of variable importance. However, random forests struggle when there are many unimportant features present. Notice that these techniques acknowledge model uncertainty only in the context of making predictions rather than investigating the models themselves.

On the other end of the spectrum from BMA, Principal Component Regression (PCR) can generate completely dissimilar models using orthogonal principal components. If we consider each principal component as a model, there is a definitive ordering to these models, where the first explains the most uncertainty and the following models explain successively less information. The models lack the same level of interpretability as models that use covariates directly.

\subsection{Model selection with uncertainty}
Generating multiple useful, but fundamentally different, models is desirable because it can provide a range of models with similar explanatory capacity that incorporate information from correlated variables. We explore a multi-model penalized regression (MMPR) algorithm that uses the SplitReg \citep{Christidis2020} objective function to generate $M$ distinct models with limited similarity. MMPR is tuned for the goal of identifying fundamentally different models to relate groups of covariates to the response rather than to reduce prediction error, as is done in \cite{Christidis2020}. The variability in these models stems from the presence of variables supplying redundant information. These models collectively capitalize on the information that correlated variables provide by including them in separate models rather than arbitrarily choosing one and excluding the others as in the single-model LASSO algorithm. These models provide multiple representations of the process under investigation, which allows subject matter experts to understand the relationship between the predictors and response from several different perspectives. These alternative explanations may generate new lines of inquiry in research that a single model interpretation would not. The SFE data analysis is a natural setting for the goals of MMPR because we want to explore several possible interpretable relationships between the correlated alloying elements and SFE.

The paper is organized in the following way. Section \ref{s:method} introduces a general form for MMPR as well as several specific cases. Section \ref{s:comp} discusses tuning and the use of coordinate descent to minimize the objective function \citep{Friedman2007,Wu2008}. Several illustrative examples using both simulated and real data are analyzed in Section \ref{s:sim}. Section \ref{sec:steel_comp} applies MMPR to the motivating fault stacking energy dataset. Finally, Section \ref{s:discussion} provides a discussion.

\section{Multi-model Penalized Regression}\label{s:method}
Suppose that we want to identify $M$ linear models that describe different relationships between the predictors and response while accounting for model uncertainty. The SplitReg framework \citep{Christidis2020} introduces an objective function that penalizes model similarity and inclusion of variables in each model. The similarity penalty encourages different subsets of or different coefficients for the covariates in the $M$ models. 

We denote the $n\times 1$ vector of responses $\bm{Y}$, the $n\times p$ standardized design matrix of covariates as $X$, and the $p \times 1$ vector of parameters for model $i \in \{ 1,\hdots,M \}$ as $\bm{\beta}_i=(\beta_{i1},\hdots,\beta_{ip})^T$. The penalized objective function is:
\begin{equation}\label{eq:obj}
  \sum_{i=1}^{M} \lVert \bm{Y}-X\bm{\beta}_i \rVert^2 + \omega \sum_{i=1}^{M-1} \sum_{j=i+1}^M P_1(\bm{\beta}_i,\bm{\beta}_j) + \lambda \sum_{i=1}^{M} P_2( \bm{\beta}_i),
\end{equation}
where $P_1(\cdot)$ is the similarity penalty, $P_2(\cdot)$ is the sparsity penalty, $\omega$ is the similarity penalty weight, and $\lambda$ is the sparsity penalty weight. The first term in the objective function is the total sum of squared errors (SSE), which is the sum of the SSEs for each of the $M$ models. This term penalizes poor model fit to the data, thereby encouraging good fits for each individual model. There is no requirement that the $M$ models contribute equally to this term. They may be ranked in terms of their individual SSE values.

The second term is the total similarity penalty, which is the sum of the similarity penalties between each pair of models. If $\omega=0$, the similarity penalty disappears and the objective function reduces to the sum of LASSO objective functions. Since these have equal weight and similarity is not penalized, the resulting models will be $M$ copies of the LASSO solution. 

Two models are similar if the same set of coefficients are present and those coefficients have comparable values. We use the similarity penalty
\begin{equation} \label{eq:similarity_pen}
P_1(\bm{\beta}_i,\bm{\beta}_j) = \sum_{k=1}^p \lvert\beta_{i k}\rvert^d \lvert\beta_{j k}\rvert^d.
\end{equation}
The absolute values ensure that the penalty is positive. They penalize the magnitude of coefficients based upon their presence in multiple models. It can be helpful to think of the similarity penalty as a shrinkage penalty applied between the models. Rather than penalizing the magnitude of a single coefficient, the term penalizes the product of the magnitudes of the same coefficients in pairs of models. The role of $d$ is to apply this penalty using different norms. Setting $d=1$ discourages the presence of a given variable $\beta_{ik}$ in multiple models $i$; it will encourage sparsity based upon similarity. This effectively allows us to choose different subsets in the different models. Setting $d=2$ encourages shrinkage of $\beta_{ik}$ in all but one model $i$ because the pairwise products are penalized, making it costly to have more than one large coefficient for the $k$th covariate. The intuitive description of the role of $d$ will be formalized in Section \ref{sec:methods_understanding}. The similarity penalty only shrinks coefficients based on the other models. With only this penalty present, i.e., $\lambda = 0$, there is no mechanism to shrink the coefficients of non-influential variables in each model.

The third term, the total sparsity penalty term, is the sum of the model sparsity penalties, which shrinks the non-influential parameters in each model to zero. We use
\begin{equation}\label{eq:sparsity_pen}
    P_2(\bm{\beta}_i) = \sum_{k=1}^p \lvert \beta_{ik} \rvert^c.
\end{equation}
The settings $c=1$ and $c=2$ correspond to LASSO and ridge penalties, respectively. One perspective of the addition of this sparsity penalty is that it reduces the variance for $\beta$ estimates at the cost of adding some bias to improve prediction \citep{Hastie2015}. For variable selection, the sparsity penalty performs includes only covariates with stronger effects.

\subsection{Understanding the Similarity Penalty}\label{sec:methods_understanding}
The full solution to Eq.~(\ref{eq:obj}) does not have a simple form. In fact, the solution is not unique, as switching the labels of the models, e.g., swapping $\bm{\beta}_1$ and $\bm{\beta}_2$, does not affect the value of the objective function. Therefore, to understand the similarity penalty, we examine the solution for one model conditioning on the others. We note that these solutions are special cases of elastic net regression~\citep{Zou2005} and thus have unique solutions for both $n\le p$ and $n>p$. 

\begin{figure}[ht]
    \centering
    \includegraphics{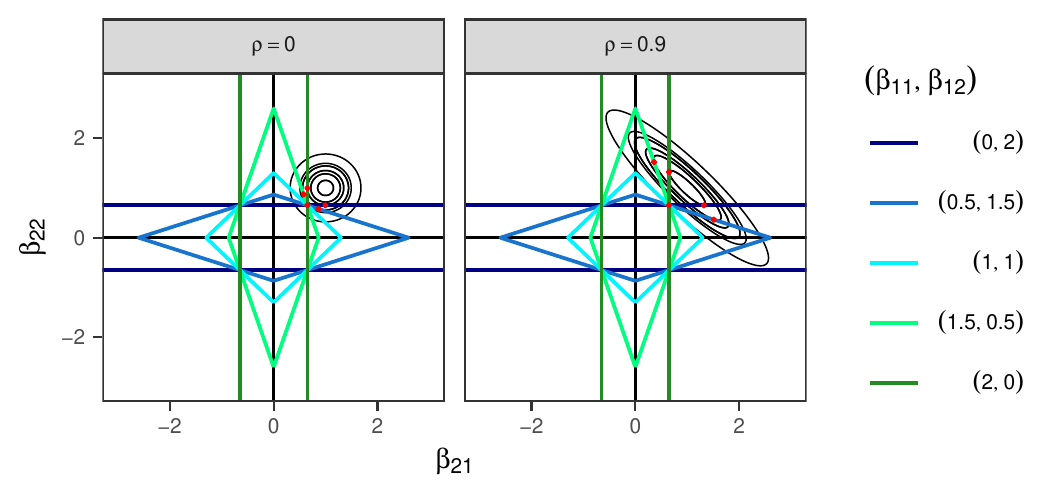}
    \caption{The similarity penalty region corresponding to $d=1$ with $\lambda=0$ for different components of Model 2, given Model 1, in the two covariate case for $M=2$ and its intersections with the SSE for $\rho=0$ (left) and $\rho=0.9$ (right).}
    \label{fig:pen_d1_contours}
\end{figure}

The parameter $d$ controls the similarity penalty. When $d=1$, the similarity penalty corresponds to an $L1$ norm, shrinking the coefficients to exactly 0. When $d=2$, the similarity penalty term is instead incorporated as a coefficient that promotes $L2$ shrinkage. The effect of $c$ similarly promotes either sparsity or shrinkage by penalizing the magnitude of the coefficients. In even a small case, with $p=2$ covariates and $M=2$ models, the similarity penalty region depends on all four components of $\hat{\bm{\beta}}$, making it difficult to visualize. Instead, we condition on Model 1 to find the penalty region with respect to Model 2 to gain a visual interpretation of the penalty. With equally important covariates whose true coefficients are $\bm{\beta}_{0} = (1,1)^T$, Figure~\ref{fig:pen_d1_contours} shows the SSE contours along with the point at which they intersect the penalty region for $d=1$ with respect to Model 2 for data generated with $\rho=0$ and $\rho=0.9$ correlation between the two covariates. When both covariates $X_1$ and $X_2$ are included with equal weight in Model 1 ($\beta_{11}=\beta_{12}=1$), they have equal penalty in Model 2. When only $X_1$ is included in Model 1 ($\beta_{11}>0$) while $X_2$ is excluded ($\beta_{12}=0$), Model 2 places a stronger penalty on the coefficient for $X_1$ than $X_2$. This effect of the penalty on the solution (red dots) is more pronounced when the covariates are highly correlated with $\rho=0.9$. 

\begin{figure}[ht]
    \centering
    \includegraphics{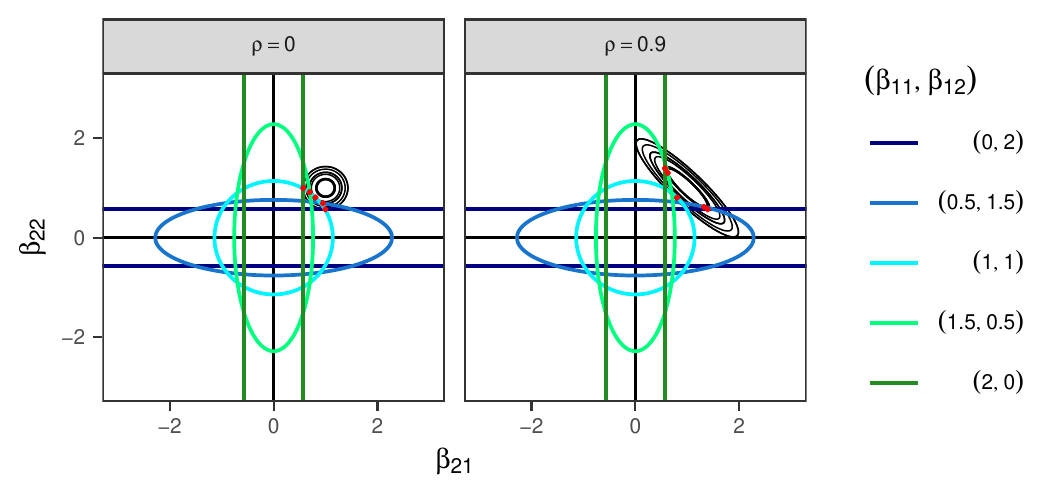}
    \caption{The similarity penalty region corresponding to $d=2$ with $\lambda=0$ for different components of Model 2, given Model 1, in the two covariate case for $M=2$ and its intersections with the SSE for $\rho=0$ (left) and $\rho=0.9$ (right).}
    \label{fig:pen_d2_contours}
\end{figure}

Figure~\ref{fig:pen_d2_contours} shows the SSE contours and penalty region for $d=2$ with respect to Model 2 for data generated with $\rho=0$ and $\rho=0.9$ correlation between the two covariates. These penalty regions mimic the penalty regions in ridge regression. Like the $d=1$ case, the penalty encourages a well-fitting model with both coefficients present in Model 2 when $\rho=0$. It encourages different models in the more highly correlated data. Similar plots for the cases in Sections~\ref{ss21}-\ref{ss24} are included in Appendix~\ref{sec:penalty}.

\subsubsection{The \texorpdfstring{$c=1$}{c1} and \texorpdfstring{$d=1$}{d1} case}\label{ss21}
Without loss of generality, we study the solution for $\bm{\beta}_1$ given $\bm{\beta}_2=\tilde{\bm{\beta}}_2,\hdots,\bm{\beta}_p=\tilde{\bm{\beta}}_p$. The solution for $\bm{\beta}_1$ when $c=1$ and $d=1$ is
\begin{equation} \label{eq:c1d1}
    \hat{\bm{\beta}}_1 = \underset{\bm{\beta}_1}{\text{argmin}} \left (\bm{Y}-X\bm{\beta}_1 \right )^T \left (\bm{Y}-X\bm{\beta}_1 \right ) + \sum_{k=1}^p \lvert \beta_{1k} \rvert \left (\lambda + \omega\sum_{j = 2}^M \lvert \tilde{\beta}_{jk} \rvert \right ).
\end{equation}
This is the adaptive LASSO objective function presented in \cite{Zou2006} with weights $w_k=\lambda+\omega\sum_{j = 2}^M \lvert \tilde{\beta}_{jk} \rvert$. The weight/penalty of $\beta_{1k}$ is large if the coefficients for covariate $k$ are absolutely large for other models. The penalty encourages model coefficients $\hat{\bm{\beta}}_1$ that do not appear first in the other $M-1$ models while shrinking those that do to $0$. Therefore, the similarity penalty encourages models to have different subsets of covariates.

\subsubsection{The \texorpdfstring{$c=1$}{c1} and \texorpdfstring{$d=2$}{d2} case}
In the case where $c=1$ and $d=2$, the solution for $\bm{\beta}_1$ is

\begin{equation} \label{eq:c1d2}
  \hat{\bm{\beta}}_1 = \underset{\bm{\beta}_1}{\text{argmin}} \left (\bm{Y}-X\bm{\beta}_1 \right )^T \left (\bm{Y}-X\bm{\beta}_1 \right ) + \omega \sum_{k=1}^{p}\left (\sum_{j=2}^{M} \tilde{\beta}_{jk}^2 \right ) \beta_{1k}^2 + \lambda \sum_{k=1}^{p} \lvert \beta_{1k} \rvert.
\end{equation}
This is the adaptive elastic net objective function presented in \cite{Ghosh2007} with constant weight $\lambda$ for the LASSO penalty and adaptive weights $\sum_{j=2}^{M} \tilde{\beta}_{jk}^2$ for the $L2$ penalty. The penalty encourages $\bm{\beta}_1$ to be a sparse model with the same coefficients that appear in the other models, but with smaller magnitudes when those coefficients are already large in another model.

\subsubsection{The \texorpdfstring{$c=2$}{c2} and \texorpdfstring{$d=1$}{d1} case}
In the case where $c=2$ and $d=1$, the solution for $\bm{\beta}_1$ is

\begin{equation} \label{eq:c2d1}
  \hat{\bm{\beta}}_1 = \underset{\bm{\beta}_1}{\text{argmin}} \left (\bm{Y}-X\bm{\beta}_1 \right )^T \left (\bm{Y}-X\bm{\beta}_1 \right ) + \lambda \sum_{k=1}^{p}  \beta_{1k}^2 + \omega \sum_{k=1}^{p}\left (\sum_{j=2}^{M} \lvert \tilde{\beta}_{jk}\rvert  \right ) \lvert \beta_{1k}\rvert.
\end{equation}
This is the adaptive elastic net objective function presented in \cite{Ghosh2007} with adaptive weights $\sum_{j=2}^{M} \lvert \tilde{\beta}_{jk}\rvert$ for the LASSO penalty and constant weight $\lambda$ for the $L2$ penalty. The penalty encourages sparsity in $\bm{\beta}_1$ when coefficients are already present in other models and shrinkage for the entire model.

\subsubsection{The \texorpdfstring{$c=2$}{c2} and \texorpdfstring{$d=2$}{d2} case}\label{ss24}
In the case where $c=2$ and $d=2$, the solution for $\bm{\beta}_1$ is

\begin{equation} \label{eq:c2d2}
  \hat{\bm{\beta}}_1 = \underset{\bm{\beta}_1}{\text{argmin}} \left (\bm{Y}-X\bm{\beta}_1 \right )^T \left (\bm{Y}-X\bm{\beta}_1 \right ) + \sum_{k=1}^{p} \left ( \lambda + \omega \sum_{j=2}^{M} \tilde{\beta}_{jk}^2 \right)\beta_{1k}^2.
\end{equation}
This is the adaptive ridge objective function as defined by \cite{Brown1980} with adaptive weights $w_k = \lambda + \omega \sum_{j=2}^{M} \tilde{\beta}_{jk}^2$ for the $L2$ penalty. This penalty encourages shrinkage for coefficients present in other models and shrinkage based on magnitudes of the coefficients.

\section{Computational Details}\label{s:comp}
We implement a coordinate descent algorithm to minimize the objective function in Eq.~(\ref{eq:obj}) and tune based on a distinct MMPR tuning metric. The \texttt{R} package \texttt{SplitReg} \citep{Christidis2020} also performs coordinate descent, but uses cross validation based on MSE to choose an appropriate sparsity and diversity penalty. The optimization algorithm used to solve MMPR cannot be done using the \texttt{SplitReg} package because the cross validation for SplitReg is based on minimizing the MSE of the group of models, rather than explicitly enforcing model diversity. The package is insufficiently flexible to extend to the different type of tuning that MMPR requires.

\subsection{Optimization}\label{s:comp:opt}
We use coordinate descent (Algorithm~\ref{alg:coord_desc}) to solve Eq.~(\ref{eq:obj}) because given all of the other coefficients, a closed form solution for the remaining coefficient is available (see Section~\ref{sec:methods_understanding}). The objective function is not convex, so it is necessary to start the algorithm at multiple starting values to find a global minimum among the local minima. The solution for the minimization of the objective function is highly dependent on starting values for the algorithm. When the number of parameters is small, it is possible to run the algorithm for an exhaustive list of starting values composed of all subsets of the covariates ($2^{Mp}$ initial values). For the values that are present, $\hat{\bm{\beta}}$ may be reasonably initialized at $\hat{\bm{\beta}}_i^{Ridge}$, $\hat{\bm{\beta}}_i^{LASSO}$, or $\hat{\bm{\beta}}_i^{OLS}$ (if it exists). \texttt{SplitReg} guarantees only a coordinatewise minimum and offers no way to specify starting values, further motivating the implementation of our own algorithm. The stopping criterion is $\left \lvert \hat{\beta}_{ik}-\tilde{\beta}_{ik} \right \rvert < \varepsilon$, where $\tilde{\beta}_{ik}$ is the value of $\hat{\beta}_{ik}$ at the previous iteration and $\varepsilon = 1e-6$. Each covariate is scaled by its $L2$ norm.

\begin{algorithm2e}[ht]
\caption{Coordinate Descent Algorithm}\label{alg:coord_desc}
\SetAlgoLined
\KwResult{Solve for $\hat{\bm{\beta}}$}
 initialize $\hat{\beta}_i = \bm{0} \quad \forall i = 1,...,M$ \;
 \While{any($\lvert\hat{\beta}_{ik} - \tilde{\beta}_{ik} \rvert > \varepsilon$)}{
 \For{$i$ in $1:M$}{
 \For{$k$ in $1:p$}{
 $\tilde{\beta}_{ik} = \hat{\beta}_{ik}$\;
 $\rho_{ik} = \sum_{l=1}^{n} x_{lk} (y_l - \sum_{h \neq k}^{p} \hat{\beta}_{ih} x_{lh})$\;
 $z_k = \sum_{l=1}^n x_{lk}^2$ \;
 $\gamma_k = (2-c)\lambda+(2-d)\omega \sum_{j \neq i}^{M} \lvert \hat{\beta}_{jk}\rvert^d$\;
 $\theta_k = (c-1)\lambda+(d-1)\omega \sum_{j \neq i}^{M} \lvert \hat{\beta}_{jk}\rvert^d$ \;
 $\hat{\beta}_{ik}=\frac{1}{\theta_k}S(\rho_{ik},\gamma_k)$\;
 }
 }
 }
\end{algorithm2e}

\subsection{Tuning}\label{s:comp:tuning}
In order to generate models, it is necessary to determine appropriate values for $M$, $\lambda$, and $\omega$. Since our motivation for a multi-model analysis is interpretation and not, say, to improve prediction, we either assume $M$ is determined by the user or examine the solutions for multiple $M$ rather than select an optimal $M$. If $M$ is chosen to be too large, the algorithm identifies the trivial model with $\bm{\beta}_i=\bm{0}$. Given $\omega$, we can generate solution paths for the model parameter estimates across $\lambda$. 

We must define how to choose an appropriate $\omega$ to control how similar the models are. There are several ways to define similarity/distance between vectors (models), which include various norms, the Pearson correlation coefficient, cosine similarity, and Tanimoto distance \citep{Han2009}. The cosine similarity is 
\begin{equation}\label{eq:cosine_sim}
d(\bm{\beta}_i,\bm{\beta}_j) = \frac{\sum_{k=1}^{n}\bm{\beta}_{ik} \bm{\beta}_{jk}}{\sqrt{\sum_{k=1}^n \bm{\beta}_{ik}^2} \sqrt{\sum_{k=1}^n \bm{\beta}_{jk}^2}},
\end{equation}
which ranges from $-1$ to $1$. Cosine similarity is particularly desirable because of its standardized values and its agreement with our qualitative definition of similarity \citep{Han2009}.

Since we are searching for different models, the cosine similarity between the $\lvert \bm{\beta}_i \rvert$ estimates across models is a natural guide for selecting $\omega$. For each value of $M$ and $\lambda$, we search for the smallest $\omega$ so that the maximum similarity between the models is less than a threshold, $\rho_{thresh}$. A high threshold will allow the models to be more similar to each other, while a lower threshold will encourage more difference between the model parameters. Using the threshold $\rho_{thresh} = 0$ works well to encourage completely different sets of parameters in models in the case where $c=1$ and $d=1$, although the lowest achievable distance depends somewhat on the data. In practice, we have found that $\rho_{thresh}=0.3$ works well to produce separate models while maintaining necessary covariates in each.

\section{Simulation}\label{s:sim}
We demonstrate MMPR on several simulated data sets to examine the behavior of the method. These simulations are presented as examples of how MMPR performs under different conditions and not to study frequentist properties of the procedure. Consequently, we provide a single simulation for each setting to illustrate the behavior of MMPR; we see consistent results for similar datasets, as shown in the appendix. In a group of highly correlated variables, a model need only include one of the group of correlated variables to get a good fit; including the others does not provide much additional information.

\subsection{Data Generation}
Each of the 7 simulated data sets is structured as blocks of correlated covariates. The 6 covariates $x_j, \; j=1,\hdots,6$ that make up $X$ have the correlation structure
\begin{equation*}
      \Gamma = I_b \otimes \Gamma_{s}^{(t)},
  \end{equation*}
where $b$ is the number of blocks of covariates, $s$ is the size of the covariate blocks, and $t$ indexes the correlation structure of the block. A block may have either a compound symmetric correlation structure, where all off-diagonal elements are $\rho$, or an AR(1) correlation structure with parameter $\rho$. For example, with block size $b=3$, the correlation structure of the blocks is
$$\Gamma_3^{(cs)} = \begin{pmatrix}
1 & \rho & \rho\\
\rho & 1 & \rho\\
\rho & \rho & 1
\end{pmatrix}, \quad \Gamma_3^{(ar)} = \begin{pmatrix}
1 & \rho & \rho^2\\
\rho & 1 & \rho\\
\rho^2 & \rho & 1
\end{pmatrix}.$$
The settings for the 7 simulated cases are detailed in Table~\ref{tab:sim_settings}. For this study we use $n=80$ data points drawn from 
$$Y \sim N(X\bm{\beta},\sigma^2 I_n),$$
where $\bm{\beta} = \begin{pmatrix}1 & 1 & 1 & 0 & 0 & 0 \end{pmatrix}^T$ and $\sigma^2=9$.

\begin{table}[ht]
    \centering
    \caption{Simulation case settings for Cases 1 through 7. The correlation between the covariates $\rho$ takes values in $\{0,0.5,0.9 \}$. Correlated blocks of block size $s \in \{2,3,6 \}$ such that there are $b \in \{3,2,1 \}$ blocks respectively make up the covariance matrix. The correlation is either compound symmetric (CS), meaning that all off diagonal elements are $\rho$, or $AR(1)$.}
    \begin{tabular}{|c|cccc|}
    \hline
        Case & $\rho$ & Block number & Block size & Correlation  \\
        \hline
        1 & 0 & 1 & 6 & -\\
        2 & 0.5 & 1 & 6 & AR(1)\\
        3 & 0.9 & 1 & 6 & AR(1)\\
        \hline
        4 & 0.5 & 2 & 3 & CS\\
        5 & 0.9 & 2 & 3 & CS\\
        \hline
        6 & 0.5 & 3 & 2 & CS\\
        7 & 0.9 & 3 & 2 & CS\\
        \hline
    \end{tabular}
    \label{tab:sim_settings}
\end{table}

\subsection{Competing Methods and Metrics}\label{s:sim:methods}
We compare our results to those obtained with Forward Stepwise Selection using the \texttt{regsubsets} function in the \texttt{leaps} package \citep{ThomasLumleybasedonFortrancodebyAlanMiller2020} and with Bayesian Model Selection using the \texttt{BMA} package \citep{Raftery2019} in \texttt{R}. For each case, a forward selection method is applied and the coefficients of the top $M=3$ models chosen based on BIC are examined as an indication of how well it captures the influential variables and produces fundamentally different models. Using a BMA analysis, the posterior probability of inclusion for each covariate is recorded as an indication of how well BMA captures influential variables in the model. Each case is also analyzed via MMPR with $c=1$ and $d=1$. The top $M=3$ forward selection models, the most likely BMA models, and the $M=3$ models generated by MMPR are compared in terms of both their coefficients and predictions. The cosine similarity in Eq.~(\ref{eq:cosine_sim}) between the absolute value of the model coefficients quantifies the similarity of the model coefficients generated by each method. The correlation of predictions generated by each pair of models within each method quantifies the prediction uncertainty stemming from model uncertainty.

\subsection{Simulation Results}\label{s:sim:results}
The solution for each of $M=3$ cases is presented as a path over a range of $\log(\lambda)$ values. As $\log(\lambda)$ increases, sparsity for the models increases. At each $\lambda$ value, $\omega$ is taken large enough that model similarity is at most 0.3. When examining the simulation plots, recall that the first three covariates (shown in green) are influential and the next three (shown in black) are not influential to the response. Each of the three covariates in the first group is denoted by a different line type; similarly for the second group of three covariates.

\begin{figure}[ht]
\centering
\includegraphics[]{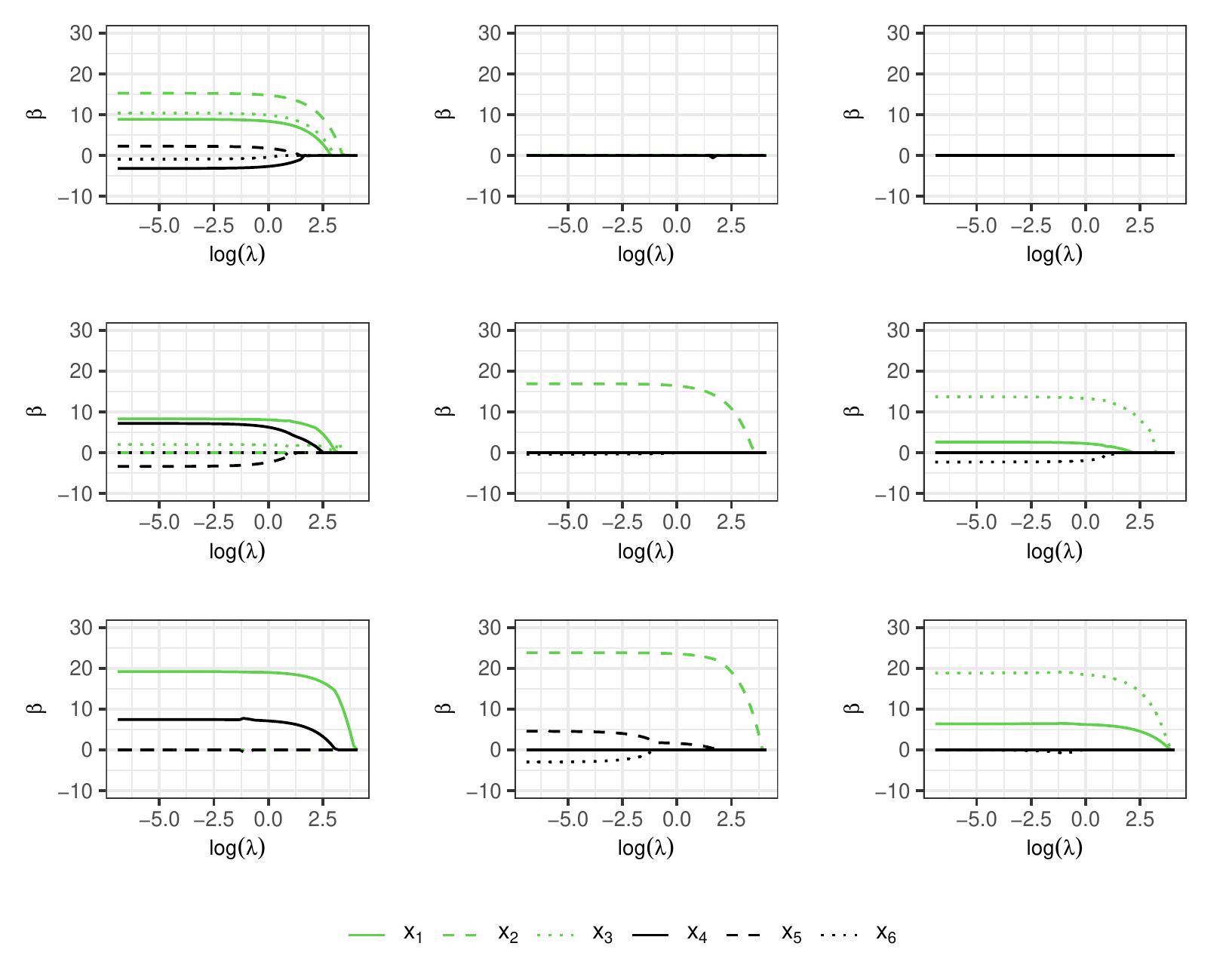}
\caption{The solution paths for MMPR for Cases 1, 2, and 3 (top to bottom), with $M=3$ models with $c=1$, $d=1$. Models are arranged horizontally within each case.}
\label{fig:sim_case123}
\end{figure}

Solution paths over $\lambda$ for the first three cases are shown in Figure~\ref{fig:sim_case123}, which demonstrates how the method behaves as autocorrelation among the covariates increases from 0 to 0.9. In Case 1, the method is unable to construct multiple useful models; one model is the best fitting model, with all three influential covariates and no non-influential ones that persist, and the others assign all coefficients to be 0 (aside from some nonzero values at high $\lambda$). This is to be expected as, in this case, all three meaningful covariates are required to explain the response. 

In Case 2, Model 1 contains $x_1$ and, to a lesser degree, $x_3$, $x_4$, and $x_5$, which do not persist for the largest values of $\lambda$. Model 2 contains $x_2$ and Model 3 contains $x_3$ and, to a lesser degree, $x_1$ and $x_6$, which persists less than the influential covariates. The noninfluential covariates in each case are penalized toward zero faster than the influential covariate as $\lambda$ increases. Note that covariates assigned to the three models tend not to be those that are most correlated; i.e., $x_1$ and $x_2$ have the maximum possible correlation and are not present in the same model. Similar behavior is observed in Case 3 except that there is more sparsity in the models, which is consistent with the increased correlation.

\begin{figure}[ht]
\centering
\includegraphics[]{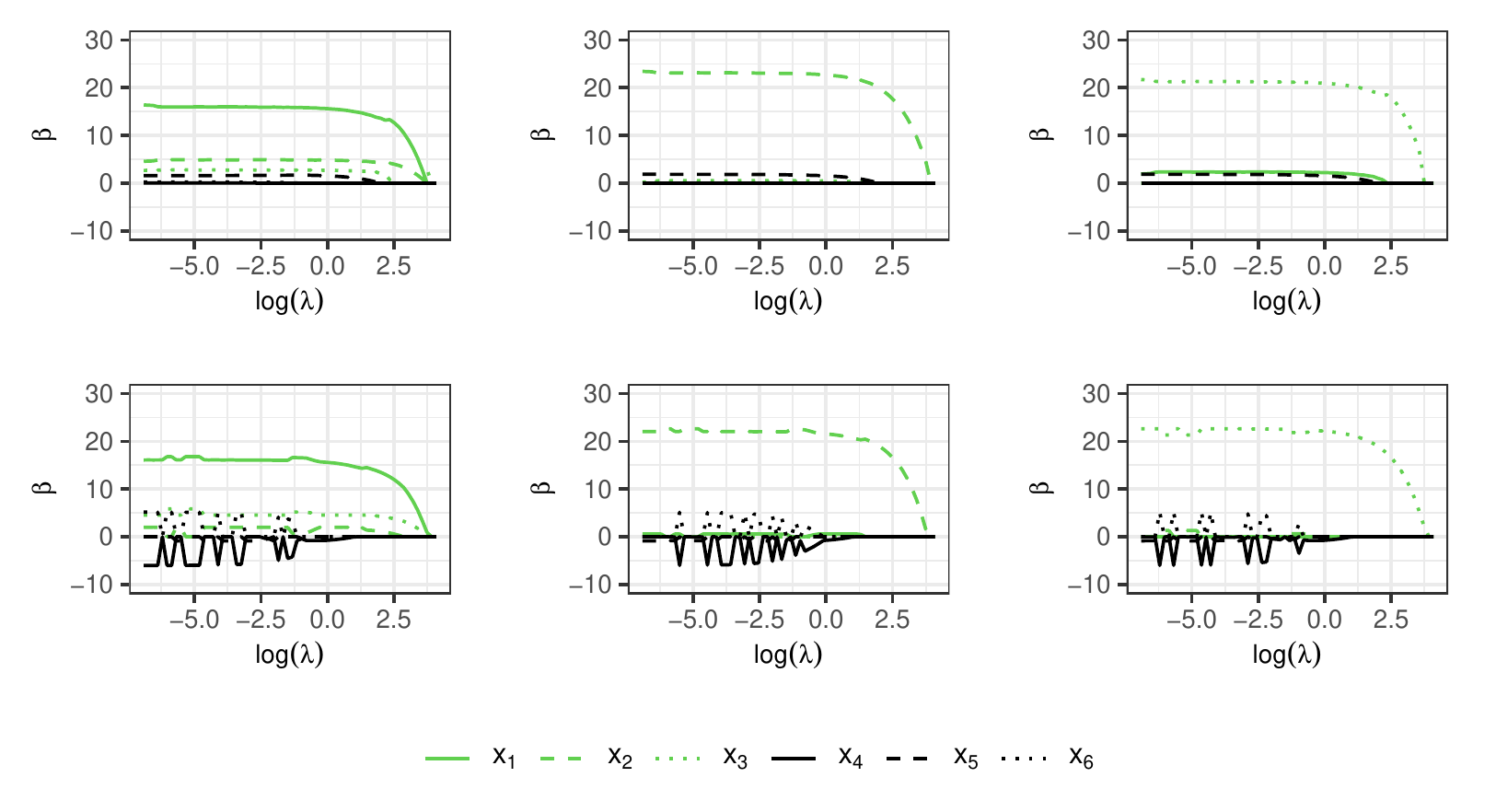}
\caption{The solution paths for MMPR for Cases 4 (top) and 5 (bottom) with $M=3$ models with $c=1$, $d=1$. Models are arranged horizontally within each case.}
\label{fig:sim_45}
\end{figure}

Figure~\ref{fig:sim_45} shows solution paths for Cases 4 and 5 and demonstrates the model selection behavior in the presence of blocks of correlated covariates whose size matches the number of models $M$. In Case 4, Model 1 contains $x_1$ at a large magnitude and $x_2$ and $x_3$ appear at a much smaller magnitude; they persist as $\lambda$ increases, while the coefficient for the noninfluential variable $x_4$ is first to disappear. Model 2 is mainly based on $x_2$, with a smaller magnitude coefficient for $x_3$. Model 3 contains $x_3$ with large magnitude and $x_1$ with smaller magnitude. Each of the three models is based mainly on a different single influential covariate while the other two influential covariates are penalized to a lower magnitude or to zero. Noninfluential covariates are quickly penalized to zero. This is to be expected because the correlation among the covariates allows each model to prioritize one of the influential covariates and leverage the correlation structure to shrink the others down to maintain dissimilarity of the models. 

In Case 5, there is a lot of switching between the noninfluential covariates in each model. This is because they are highly correlated. For this case, Model 1 contains $x_1$ with large magnitude and $x_3$ with smaller magnitude; Model 2 contains $x_2$ with large magnitude and $x_1$ with smaller magnitude; and Model 3 contains $x_3$ with large magnitude. Each model contains a single influential covariate with large magnitude. This is to be expected because the high correlation allows each model to select only one or two of the influential covariates, leveraging the strong correlation to include a smaller number of variables.

\begin{figure}[ht]
\centering
\includegraphics[]{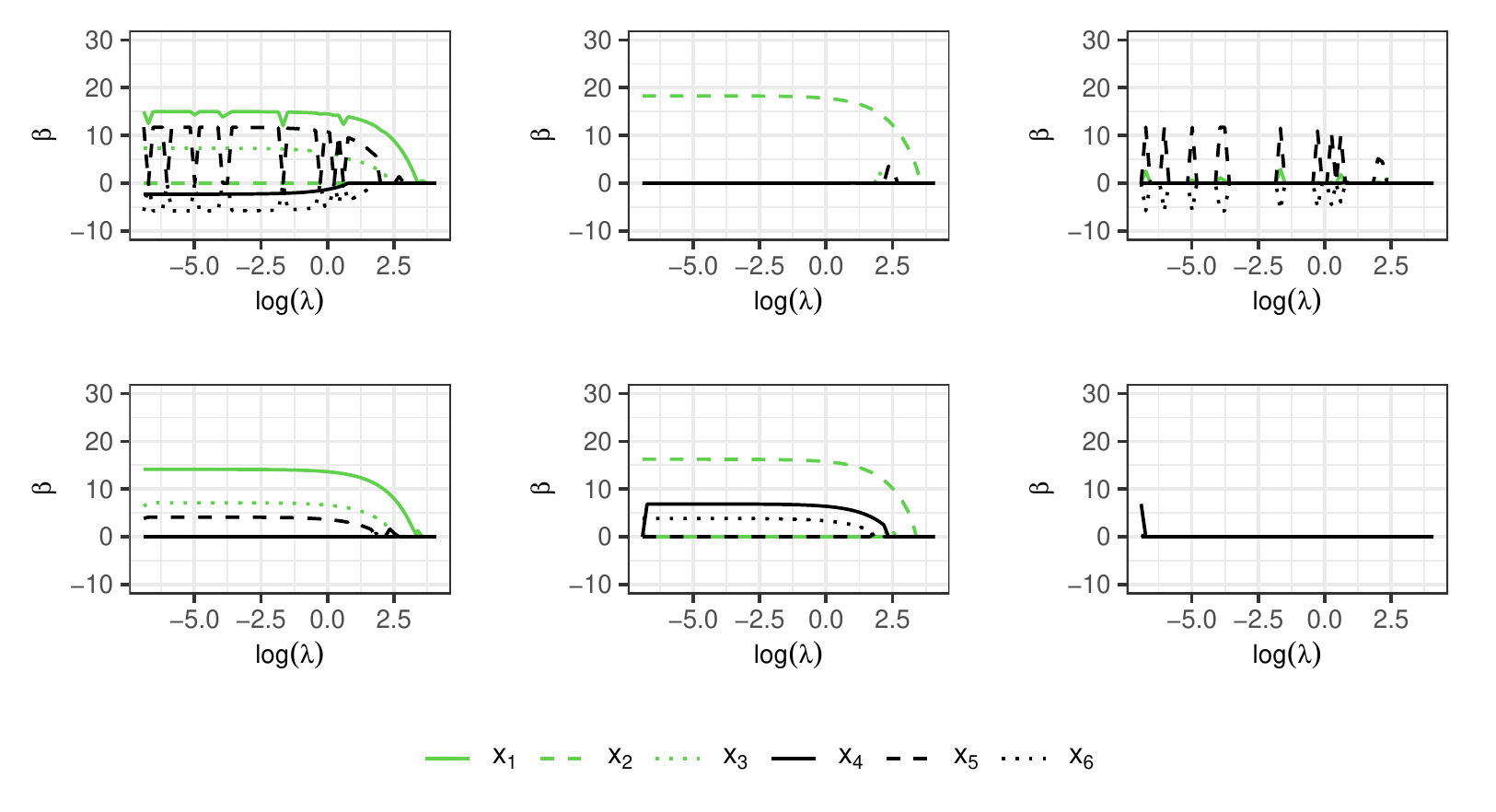}
\caption{The solution paths for MMPR for Cases 6 (top) and 7 (bottom) with $M=3$ models with $c=1$, $d=1$. Models are arranged horizontally within each case.}
\label{fig:sim_case67}
\end{figure}

The solution paths for Cases 6 and 7 in Figure~\ref{fig:sim_case67} demonstrate the model selection behavior when the number of models is greater than the size of the blocks of correlated covariates. In Case 6, Model 1 contains $x_1$ and $x_3$ while Model 2 contains $x_2$ and Model 3 is essentially empty. The models split the correlated covariates to achieve two models each containing uncorrelated covariates. The method automatically accounts for the mismatch in block size and number of models by reducing the third model to noise and eventually zero. The results are similar for Case 7, but the magnitudes of the coefficients are larger.

\subsection{Comparison with Competing Methods}
Forward selection and BMA were used to analyze each simulation case presented in Table~\ref{tab:sim_settings}. In particular, Case 4 contains groups of correlated variables. With this type of correlated data, BMA can have trouble assigning high probability to all influential variables; prediction leans on the high correlation of all of the influential variables. However, the correlation is still low enough that there is not much model uncertainty associated with BMA.

Examining Case 4 in depth, we compare the similarity among the models generated by each method. The MMPR model is discussed with $\lambda=14.8$, based on the $\lambda$ suggested by cross validation for LASSO. Forward selection identifies nested models that are not very different from each other as shown in Table~\ref{tab:fs_coeffs}. BMA identifies several models that are mostly similar to each other. MMPR identifies three sparse models, each based primarily on a single influential covariate.

\begin{figure}[ht]
\centering
\includegraphics[width=\linewidth, trim=0 60 0 40, clip]{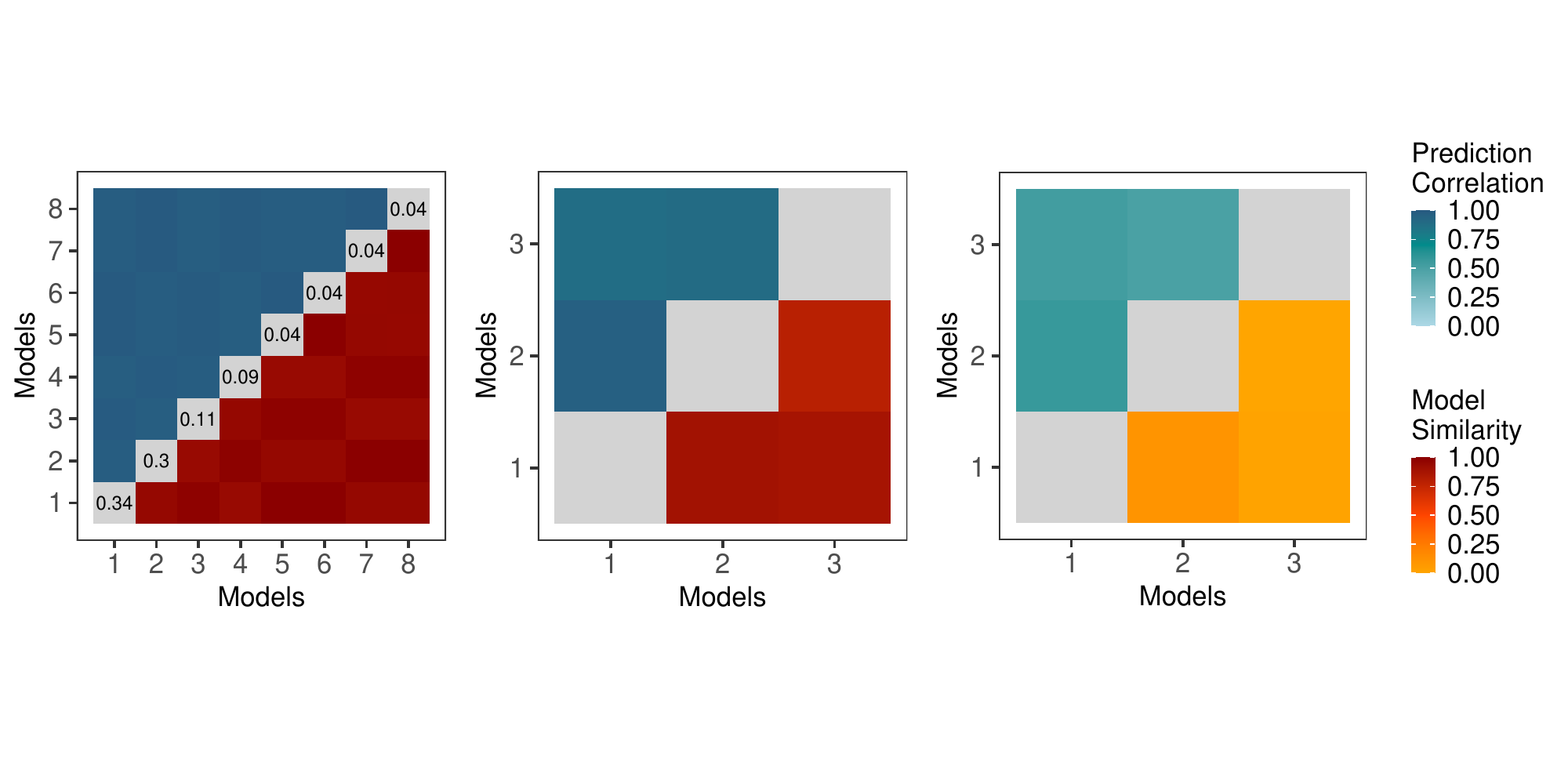}
\caption{Comparison of the model coefficient cosine similarity and prediction correlations of the BMA (left), forward selection (center), and MMPR (right) models for Case 4. The horizontal and vertical axes correspond to the models. The upper left corner measures the pairwise correlation between the model predictions ($X\bm{\beta}_i$) while the lower right corner shows similarity between the models $\bm{\beta}_i$. The diagonal displays the posterior probability in the BMA case.}
\label{fig:sim1_cor}
\end{figure}

Figure~\ref{fig:sim1_cor} shows that the models for the MMPR method are dissimilar while the forward selection and BMA models are highly similar, notably among the models assigned the highest probabilities for BMA. The correlation between posterior predictions in MMPR is lower than for BMA and forward selection. Therefore, with MMPR, we achieve uncorrelated models that still contain the influential variables.

Stepping back again to consider all simulation scenarios, the coefficients for the top 3 forward selection models are shown in Table~\ref{tab:fs_coeffs}. While the influential variables are included in the each of the top 3 models for all cases, the models are not very dissimilar from each other in cases with high correlation, such as Cases 3, 5, and 7, failing to provide the insight that perhaps the model need only be based on one of the correlated covariates.

\begin{table}[ht]
\centering
\caption{Forward selection coefficient estimates of top M models for simulation data} 
\label{tab:fs_coeffs}
\begin{tabular}{|c|cccccc|}
  \hline
 & x1 & x2 & x3 & x4 & x5 & x6 \\ 
  \hline
 & 10.40 & 8.88 & 15.28 & - & - & - \\ 
  Case 1 & 10.40 & 8.88 & 15.28 & -3.18 & - & - \\ 
    & 10.40 & 8.88 & 15.28 & -3.18 & 2.27 & - \\ 
   \hline
   & 10.40 & 12.32 & 9.16 & - & - & - \\ 
  Case 2 & 9.16 & 3.53 & 10.56 & -3.18 & - & - \\ 
      & 10.40 & 16.90 & 15.28 & -3.18 & 2.27 & - \\ 
   \hline
     & 26.15 & 12.32 & 9.16 & - & - & - \\ 
  Case 3 & 19.53 & 3.53 & 8.18 & -3.18 & - & - \\ 
        & 19.53 & 16.90 & 9.30 & -3.18 & 2.27 & -1.54 \\ 
   \hline
       & 12.77 & 6.50 & 16.20 & - & - & - \\ 
  Case 4 & 19.53 & 18.37 & 14.93 & -3.18 & - & - \\ 
          & 12.77 & 6.50 & 16.20 & -3.18 & 3.57 & -1.54 \\ 
   \hline
         & 12.77 & 6.50 & 23.10 & - & - & - \\ 
  Case 5 & 19.53 & 13.64 & 10.51 & -3.18 & - & - \\ 
            & 12.77 & 13.64 & 10.51 & -1.40 & 3.57 & -1.54 \\ 
   \hline
           & 6.18 & 7.82 & 8.83 & - & 14.40 & - \\ 
  Case 6 & 8.83 & 7.82 & 10.51 & -3.18 & 14.40 & - \\ 
              & 6.18 & 7.82 & 11.74 & -1.40 & 14.40 & -5.81 \\ 
   \hline
             & 6.18 & 16.24 & 7.13 & - & 14.40 & - \\ 
  Case 7 & 8.83 & 16.24 & 10.51 & -3.18 & 14.40 & - \\ 
                    & 6.18 & 16.24 & 7.13 & -1.40 & 4.12 & -5.81 \\ 
   \hline
\end{tabular}
\end{table}

For BMA, the corresponding posterior probabilities of the coefficients for each case are shown in Table~\ref{tab:bma_prob}. In Case 1, the probabilities for inclusion of each influential parameter are 100\% and for each noninfluential parameter are low. BMA performs particularly well in the presence of independent covariates. BMA does not convincingly identify all three of the influential covariates in cases where correlation is high, notably in Cases 3, 5, and 7. While this does not affect prediction, it does provide an inaccurate interpretation of the influential variables.

The forward selection method is unable to create very different models, while MMPR is. As a side effect of creating the disparate models, MMPR is able to identify each of the three influential covariates in all cases. In contrast, for BMA, the posterior probability of inclusion for the influential covariates is often low in the presence of correlation. BMA focuses on prediction, which can lead to these low inclusion probabilities for highly correlated influential covariates, sacrificing interpretation of variable selection.

\begin{table}[ht]
\centering
\caption{BMA posterior probabilities of covariates in simulations} 
\label{tab:bma_prob}
\begin{tabular}{|c|cccccc|}
  \hline
 & x1 & x2 & x3 & x4 & x5 & x6 \\ 
  \hline
Case 1 & 1.00 & 1.00 & 1.00 & 0.16 & 0.11 & 0.08 \\ 
  Case 2 & 0.11 & 1.00 & 0.89 & 0.06 & 0.09 & 0.09 \\ 
  Case 3 & 0.95 & 0.16 & 0.15 & 0.09 & 0.07 & 0.06 \\ 
  Case 4 & 0.54 & 1.00 & 1.00 & 0.08 & 0.21 & 0.08 \\ 
  Case 5 & 0.13 & 0.43 & 0.83 & 0.07 & 0.06 & 0.06 \\ 
  Case 6 & 0.65 & 1.00 & 0.54 & 0.10 & 0.94 & 0.32 \\ 
  Case 7 & 0.07 & 1.00 & 0.44 & 0.36 & 0.17 & 0.15 \\ 
   \hline
\end{tabular}
\end{table}

\section{Stacking Fault Energy from Steel Composition}\label{sec:steel_comp}
In this section, we apply MMPR to the steel alloy composition data from Section \ref{s:intro}. First, consider the correlation structure present in the data (Figure~\ref{fig:cor_steel_comp}). Alloys are fabricated, so the correlation structure depends on the choices made in manufacturing. A quirk of composition data is that, since composition must sum to $100\%$, the fraction of one element must decrease as another increases. Consequently, it may be challenging to determine whether a change in SFE is due to the decrease of one element or the increase in another. There are three pairs of elements with moderately strong negative correlation (magnitude greater than $0.7$): Fe and Ni, Mn and Cr, C and Cr. Many of the other additive compositions are so small compared to elements like Ni and Cr that they do not have strong correlations with Fe. Elements with moderately strong positive correlations are C and Mn and Ni and Cr. Carbon (C) and manganese (Mn) occur together in many alloys because, at relatively low concentrations, the presence of Mn helps a heated steel alloy absorb C during a manufacturing step called carburization, improving the hardness of the material \citep{Mg}. Cr is soluble in Ni and, at certain concentrations, the pair is known to improve properties like corrosion/wear resistance and hardness \citep{AZoM6}. 

\begin{figure}[ht]
\centering
\includegraphics[]{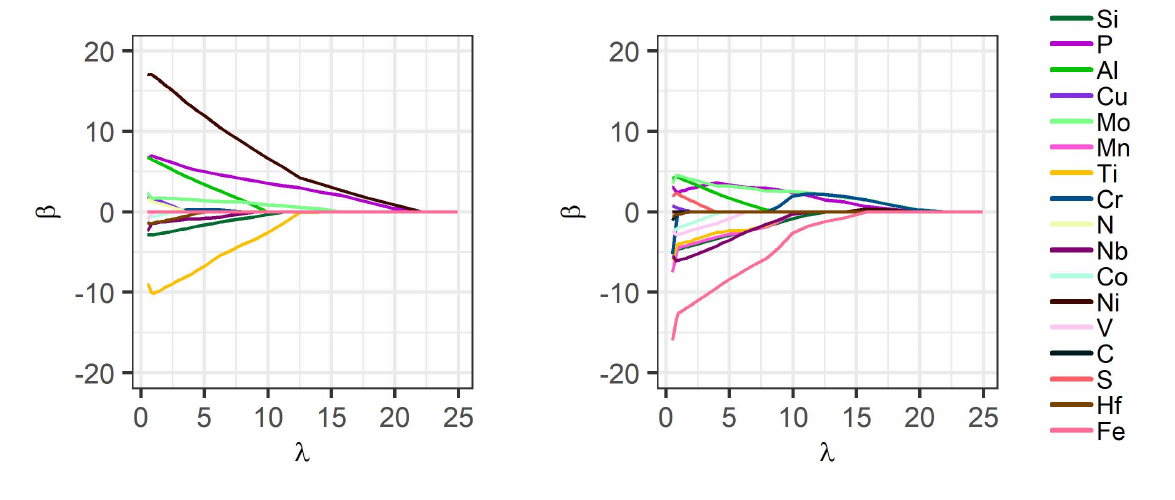}
\caption{The solution paths for MMPR analysis of stacking fault energy data with $M=2$ models at settings $c=1$, $d=1$.}
\label{fig:sfe_paths_M2}
\end{figure}

An MMPR analysis was conducted on the SFE dataset with $c=1$ to encourage sparsity of the models and $d=1$ to encourage sparse dissimilarity between $M=2$ models. Solutions were calculated for a sequence of $\lambda$ values with $\omega$ chosen to limit model similarity to at most $0.3$. Figure~\ref{fig:sfe_paths_M2} shows the solution paths for the two models generated by MMPR. The MMPR results include elements with low correlation, Si, P, Al, Cu, Mo, Ti, Nb, Co, V, and Hf, in both models at low values of $\lambda$, where sparsity is due to similarity of models. The information provided by these elements is only possible to incorporate by including each one. The elements Ni, P, Mo, and Ti persist in Model 1 while Fe, P, Cr, Mo, persist in Model 2. Recall that Ni and Fe as well as Ni and Cr are correlated. Each model identifies one of these elements for most values of $\lambda$, although Ni is shared to some degree at values $\lambda \approx 15$. Only one element in the correlated pair is necessary in each model because they provide redundant information. 

For the less sparse models, with low $\lambda$, it is interesting to note that Mn and C appear together with negative coefficients. Even though they are correlated, the information provided is not redundant and their appearance together despite their correlation may be compared to the existence of an interaction term. This is consistent with the idea that the inclusion of these two elements as a pair impacts steel properties and further extends this pattern to SFE. The predictor elements P and Mo are relatively uncorrelated with each other and other predictors and appear in both models. 

While Fe must by definition be present in the alloy, it is interesting conceptually to consider both models. Does SFE increase due to large values of Ni or small values of Fe? In context, it may be more enlightening to think of change in SFE in terms of the increasing Ni additive composition (Model 1) rather than the decreasing iron composition (Model 2). MMPR separates out these two perspectives.

\begin{figure}[ht]
\centering
\includegraphics[]{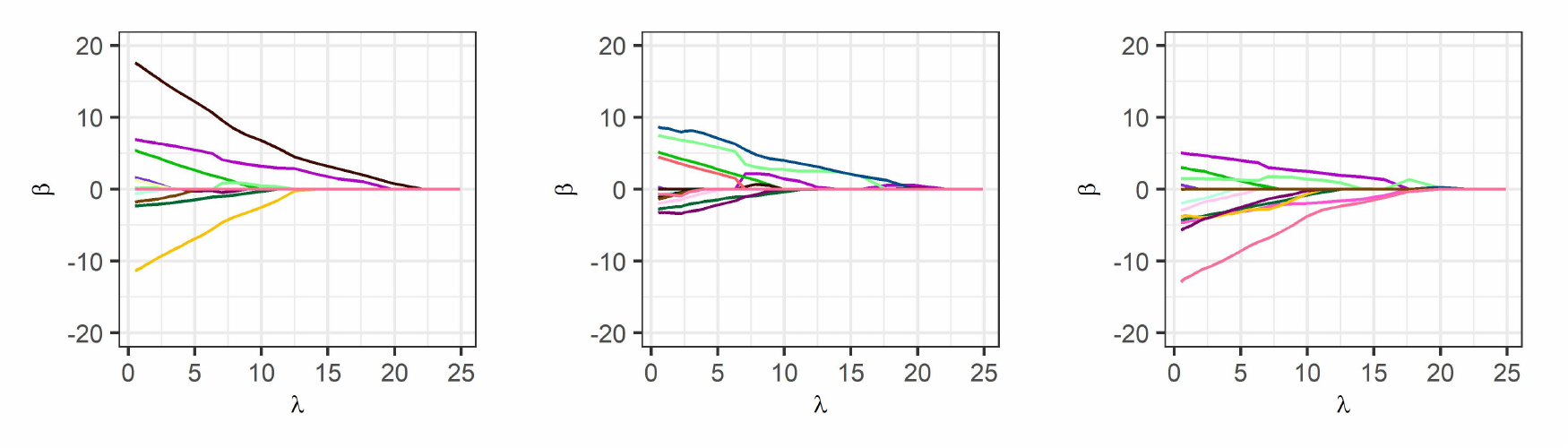}
\caption{The solution paths for MMPR analysis of stacking fault energy data with $M=3$ models at settings $c=1$, $d=1$.}
\label{fig:sfe_paths_M3}
\end{figure}

A second MMPR analysis with $M=3$ models and similarity limited to $0.5$ shown in Figure~\ref{fig:sfe_paths_M3} offers complementary insights. Again, some elements appear in all three models: Si, Al, Cu, Mo, Co, V, and Hf. As in the $M=2$ case, Ni and Fe do not appear together in the same model. Unlike in the $M=2$ case, at low $\lambda$, C and Mn appear in separate models; the opportunity to fit different models overrides the benefit of these elements appearing together. 

\begin{figure}[ht]
    \centering
    \includegraphics{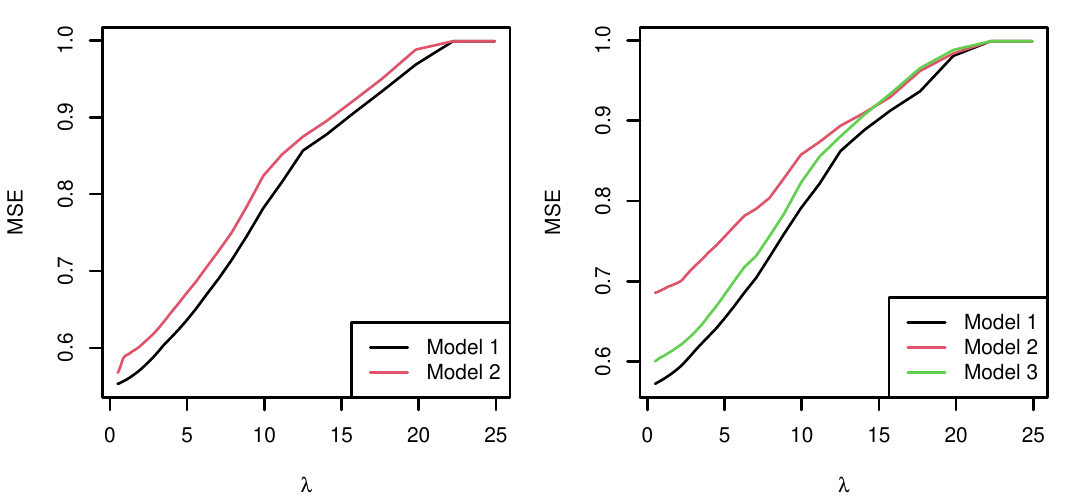}
    \caption{MSE for SFE models with $M=2$ (left) and $M=3$ (right) models. }
    \label{fig:sfe_msel}
\end{figure}

Mean squared error (MSE) is used to measure the quality of each of the $M$ models as shown in Figure~\ref{fig:sfe_msel}. While similar, the models do not have exactly the same values for MSE.

\citeauthor{Chaudhary2017} focus mainly on classification based on black box algorithms, while this MMPR analysis yields interpretable results relating composition by element to SFE. MMPR is able to identify paths for two dissimilar models to explain stacking fault energy as a function of composition. This information is best viewed in the context of the materials science application to learn about the associations between composition and SFE. The models suggest that the presence of P, Al, Cu, Mo, N, Ni, and S are positively associated with SFE and Si, Mn, Ti, Cr, Nb, Co, V, C, Hf, and Fe are negatively associated with SFE. It suggests the importance of the interaction between Mn and C on SFE and presents two different views of how Fe versus Ni impacts SFE. These intuitions enrich the current understanding of the impact of composition on SFE and can help intuitively guide composition proposals to achieve either high or low SFE.  

\section{Discussion}\label{s:discussion}
MMPR adapts the SplitReg \citep{Christidis2020} framework to perform variable selection with model uncertainty. Tuning parameter selection is intentionally designed to emphasize model dissimilarity, instead of prediction, to achieve several different explanations of the data. Different settings for the proposed objective function encourage shrinkage or sparsity behavior in terms of model similarity and coefficient magnitude. Sparse settings with $c=1$ and $d=1$ perform variable selection in terms of both magnitude and similarity. MMPR tends to allocate correlated variables to different models, so that, collectively, the models contain all of the influential variables. This is an improvement over LASSO, which tends to include a single variable in a correlated group arbitrarily based on the sample. With respect to model uncertainty, MMPR chooses dissimilar models as opposed to Bayesian model averaging, where the most probable models are frequently similar to each other. MMPR also lends itself to interpretable models for the response.

MMPR is performed on several simulated cases to demonstrate the behavior of the method on the variables selected in the individual models. When covariates are correlated, they provide redundant information, so each of the multiple models need only include one or two covariates to capture the information. Collectively, the multiple models identify all influential covariates, where BMA or LASSO might deem some unimportant. It tends to include correlated variables in separate models and uncorrelated variables in all models, identifying models with different covariates present when possible. We then apply the approach to steel alloy composition data to model SFE and generate two different interpretations of a composition model for SFE. These multiple views, in the context of materials expertise, highlight two perspectives for relating alloy composition to SFE.

Future work concerned with inserting physically accurate expert knowledge regarding the presence of certain variables in all possible models could be conducted by modifying the objective function in Eq. (\ref{eq:obj}). It is possible to facilitate the inclusion of specific variables in all models by removing the similarity penalty for the associated covariates, but further study of this modification is open for investigation. While this paper only explores MMPR for linear models, the approach may be extended to generalized linear models as in ~\cite{Hastie2015} by minimizing the negative log-likelihood along with the sparsity and similarity penalties. This greatly increases the number of problems for which MMPR can be applied.


\section*{Funding}
This material is based upon work supported by the National Science Foundation under grant DGE-1633587, CMMT-1844484, and CMMT-2022254. 

\subsection*{Competing interests}
No potential competing interest was reported by the authors.

\begin{singlespace}

\end{singlespace}

\FloatBarrier

\appendix

\section{MMPR penalty regions}\label{sec:penalty}
The penalty regions for the cases where the non-similarity related sparsity/shrinkage term is present ($\lambda \neq 0$) are modified from the regions shown in Figure~\ref{fig:pen_d1_contours} and Figure~\ref{fig:pen_d2_contours}. The main difference is that the introduction of the general sparsity term with $\lambda \neq 0$ bounds all of the covariates; even when $\beta_{11}=0$, $\beta_{21}$ is constrained. They combine the linear behavior of the absolute value portion of the penalty with the curved behavior of the squared portion of the penalty.

\begin{figure}[h]
    \centering
    \includegraphics{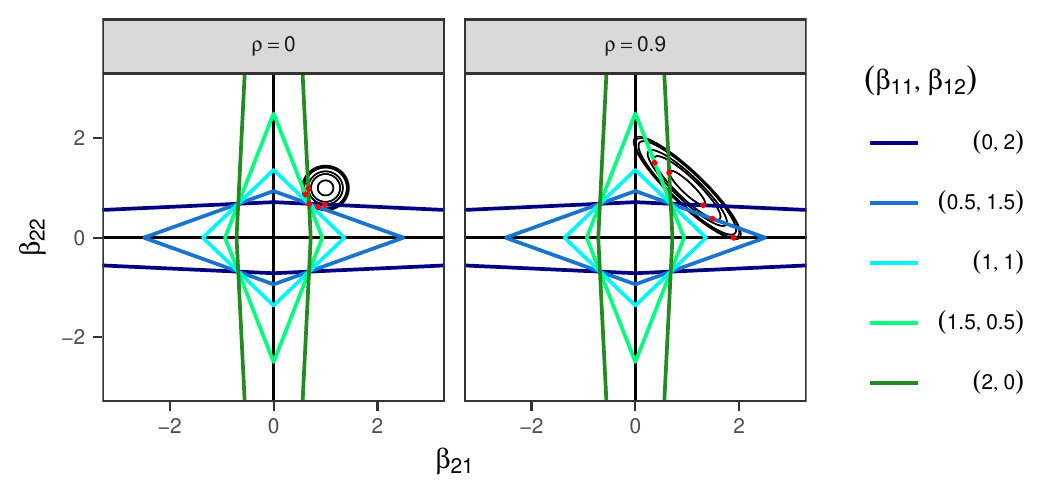}
    \caption{The similarity penalty region corresponding to settings $c=1$, $d=1$ for different components of Model 2, given Model 1, in the two covariate case for $M=2$ and its intersections with the SSE for $\rho=0$ (left) and $\rho=0.9$ (right).}
    \label{fig:pen_c1d1_contours}
\end{figure}

\begin{figure}[h]
    \centering
    \includegraphics{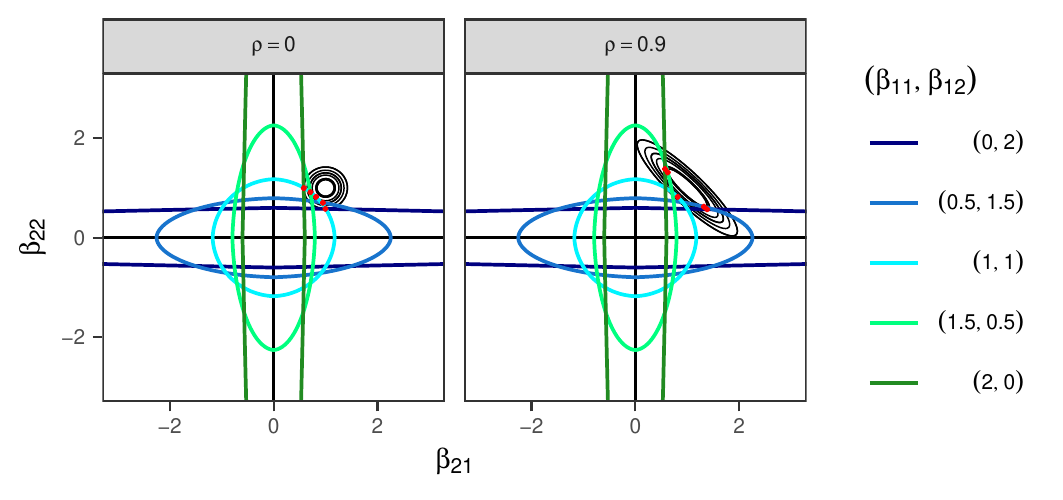}
    \caption{The similarity penalty region corresponding to settings $c=1$, $d=2$ for different components of Model 2, given Model 1, in the two covariate case for $M=2$ and its intersections with the SSE for $\rho=0$ (left) and $\rho=0.9$ (right).}
    \label{fig:pen_c1d2_contours}
\end{figure}

\begin{figure}[h]
    \centering
    \includegraphics{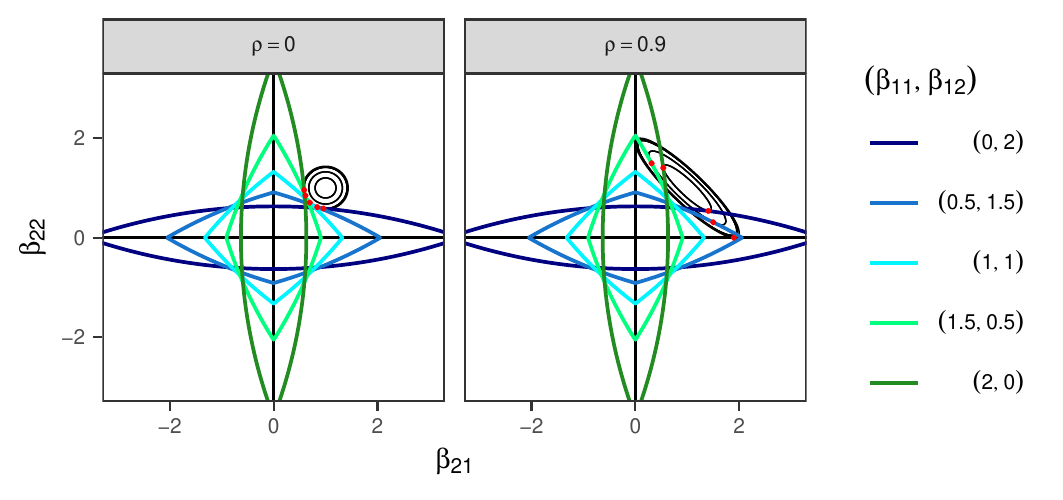}
    \caption{The similarity penalty region corresponding to settings $c=2$, $d=1$ for different components of Model 2, given Model 1, in the two covariate case for $M=2$ and its intersections with the SSE for $\rho=0$ (left) and $\rho=0.9$ (right).}
    \label{fig:pen_c2d1_contours}
\end{figure}

\begin{figure}[h]
    \centering
    \includegraphics{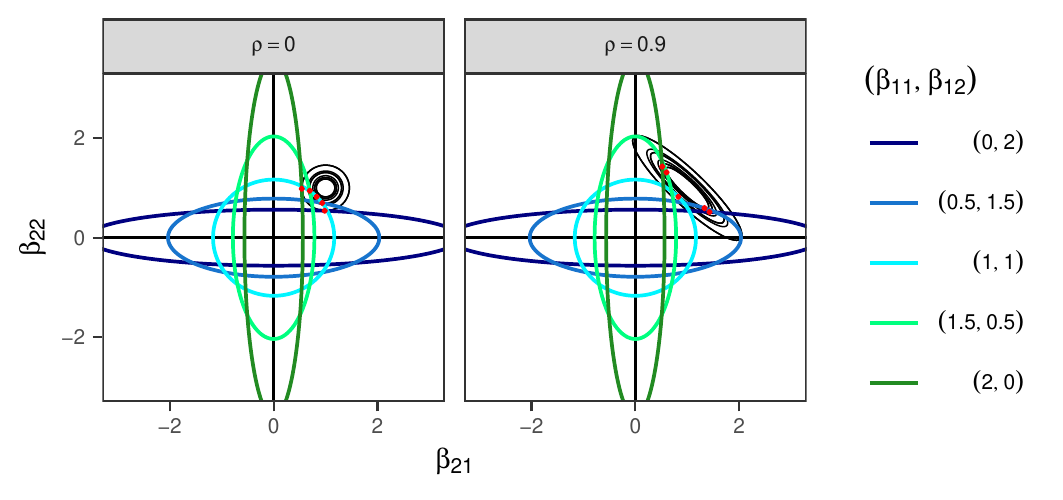}
    \caption{The similarity penalty region corresponding to settings $c=2$, $d=2$ for different components of Model 2, given Model 1, in the two covariate case for $M=2$ and its intersections with the SSE for $\rho=0$ (left) and $\rho=0.9$ (right).}
    \label{fig:pen_c2d2_contours}
\end{figure}

\FloatBarrier
\section{Multiple simulation datasets}
In Section~\ref{s:sim}, a single dataset from each simulation case was analyzed in detail. The following results for several datasets are included to confirm that the analysis behavior demonstrated on the single datasets extends to other datasets with the same data generation process. MMPR was run on 16 simulated datasets generated by each case. Each model is calculated at the $\lambda$ value chosen by cross-validation for the LASSO fit. Table~\ref{nonzero} tabulates the proportion of samples in which each covariate is present in each model. For every case, similar models are generated over the 16 sample datasets in terms of inclusion of the influential variables (though not necessarily the non-influential variables). The covariate inclusion over multiple datasets indicates that the single result cases shown in the paper are representative of MMPR analyses for these simulation cases in general.

\begin{table}[ht]
\centering
\caption{Proportion of nonzero coefficient estimates in each simulation case.} 
\label{nonzero}
\begin{tabular}{|c|c|cccccc|}
  \hline
Case & Model & x1 & x2 & x3 & x4 & x5 & x6 \\ 
  \hline
 & Model 1 & 1.00 & 1.00 & 1.00 & 0.88 & 0.94 & 1.00 \\ 
  1 & Model 2 & 0.00 & 0.00 & 0.00 & 0.12 & 0.06 & 0.00 \\ 
   & Model 3 & 0.00 & 0.00 & 0.00 & 0.00 & 0.00 & 0.00 \\ 
   \hline
 & Model 1 & 1.00 & 0.00 & 1.00 & 1.00 & 0.00 & 0.56 \\ 
  2 & Model 2 & 0.00 & 1.00 & 0.00 & 0.00 & 1.00 & 0.50 \\ 
   & Model 3 & 1.00 & 0.00 & 1.00 & 0.00 & 0.00 & 0.56 \\ 
   \hline
 & Model 1 & 1.00 & 0.00 & 0.00 & 1.00 & 0.00 & 0.44 \\ 
  3 & Model 2 & 0.00 & 1.00 & 0.00 & 0.00 & 1.00 & 0.69 \\ 
   & Model 3 & 1.00 & 0.00 & 1.00 & 0.00 & 0.00 & 0.56 \\ 
   \hline
 & Model 1 & 1.00 & 1.00 & 1.00 & 0.50 & 0.50 & 0.56 \\ 
  4 & Model 2 & 1.00 & 1.00 & 1.00 & 0.44 & 0.62 & 0.62 \\ 
   & Model 3 & 1.00 & 1.00 & 1.00 & 0.38 & 0.88 & 0.50 \\ 
   \hline
 & Model 1 & 1.00 & 1.00 & 0.00 & 0.44 & 0.69 & 0.38 \\ 
  5 & Model 2 & 0.00 & 1.00 & 1.00 & 0.50 & 0.81 & 0.69 \\ 
   & Model 3 & 1.00 & 0.00 & 1.00 & 0.56 & 0.62 & 0.31 \\ 
   \hline
 & Model 1 & 1.00 & 0.00 & 1.00 & 0.00 & 0.38 & 0.56 \\ 
  6 & Model 2 & 0.00 & 1.00 & 0.00 & 1.00 & 0.50 & 0.25 \\ 
   & Model 3 & 0.12 & 0.00 & 0.00 & 0.00 & 0.12 & 0.19 \\ 
   \hline
 & Model 1 & 1.00 & 0.00 & 1.00 & 0.00 & 0.69 & 0.38 \\ 
  7 & Model 2 & 0.00 & 1.00 & 0.00 & 1.00 & 0.31 & 0.56 \\ 
   & Model 3 & 0.00 & 0.00 & 0.00 & 0.00 & 0.06 & 0.06 \\ 
   \hline
\end{tabular}
\end{table}

\end{document}